\documentclass{PoS}

\title{Phase structure of topological insulators by lattice strong-coupling expansion}

\ShortTitle{Phase structure of topological insulators by lattice strong-coupling expansion}

\author{\speaker{Yasufumi Araki}\\
        Department of Physics, The University of Texas at Austin, Austin, TX 78712, USA\\
        E-mail: \email{y.araki85@gmail.com}}

\author{Taro Kimura\\
		Institut de Physique Th\'eorique, CEA, IPhT, F-91191 Gif-sur-Yvette, France\\
        Mathematical Physics Laboratory, RIKEN Nishina Center, Saitama 351-0198, Japan}
\author{Akihiko Sekine, Kentaro Nomura\\
        Institute for Materials Research, Tohoku University, Sendai 980-8577, Japan}
\author{Takashi Z.~Nakano\\
		Kozo Keikaku Engineering Inc., Tokyo 164-0012, Japan}

\abstract{
The effect of the strong electron correlation on the topological phase structure of 2-dimensional (2D) and 3D
topological insulators is investigated, in terms of lattice gauge theory.
The effective model for noninteracting system is constructed similarly to the lattice fermions with the Wilson term,
corresponding to the spin-orbit coupling.
Introducing the electron-electron interaction as the coupling to the gauge field,
we analyze the behavior of emergent orders by the strong coupling expansion methods.
We show that there appears a new phase with the in-plane antiferromagnetic order in the 2D topological insulator,
which is similar to the so-called ``Aoki phase'' in lattice QCD with Wilson fermions.
In the 3D case, on the other hand, there does not appear such a new phase,
and the electron correlation results in the shift of the phase boundary between the topological phase and the normal phase.}

\FullConference{31st International Symposium on Lattice Field Theory - LATTICE 2013\\
		July 29 - August 3, 2013\\
		Mainz, Germany}

\newcommand \beq{\begin{eqnarray}}
\newcommand \eeq{\end{eqnarray}}

\newcommand{\bfr}{\mathbf{r}}

\newcommand{\bfk}{\mathbf{k}}
\newcommand{\bfK}{\mathbf{K}}

\newcommand{\vf}{v_{_F}}

\begin{document}

\section{Introduction}
Topological insulator is the new class of material that is attracting a great interest in the field of materials physics over the decade \cite{Hasan-Kane,Qi-Zhang}.
It is characterized by the gapless surface (edge) states even though the bulk has a finite bandgap,
which are protected against any perturbations or disorder respecting the symmetry of the system.
Emergence of such kind of surface (edge) states is ensured by the nontrivial topology of the wavefunction of electrons,
which is characterized by the topological invariants, such as the winding number.

The effect of electron correlation on the band structure has always been an important problem in the materials physics.
For instance, even in the nontopological systems, such as graphene,
there has been a proposal that the electron correlation can drive the system toward the excitonic instability,
turning the system from semimetal into a Mott insulator \cite{Physics_2009}.
As the electron-electron interaction is mediated by a gauge field, namely the electromagnetic field,
this scenario is similar to the spontaneous chiral symmetry breaking and the consequent quark mass generation mechanism in quantum chromodynamics (QCD).
One of the authors investigated this mechanism in graphene by the strong coupling expansion technique of lattice gauge theory \cite{Araki_2010},
which has been one of the methods to analyze the QCD phase structure qualitatively \cite{Kawamoto_Smit_1981,TZN_2010}.
Although it is suggested that the interaction strength in the realistic graphene is not strong enough for the excitonic instability \cite{Ulybyshev_2013,Drut_2013},
the importance of this scenario is unchanged in the context of other strongly correlated systems, such as the ultracold fermionic atoms on the optical lattice.

In this paper, we focus on such an effect of electron correlation on the phase structure of topological insulators,
described by the boundary between the topological insulator and the normal insulator phases.
We formulate the models of topological insulators in terms of massive lattice femions:
in 2-dimensions (2D), we take the Kane--Mel\'e model for quantum spin Hall (QSH) insulator, one of the classes of topological insulators in 2D,
which is a straightforward extension of the conventional tight-binding model of graphene.
In 3D, we make use of the Wilson fermion on the hypothetical square lattice.
As in the previous studies on graphene,
we incorporate the electron-electron interaction mediated by the electromagnetic field in terms of lattice gauge theory,
and analyze the emergent orders by the strong coupling expansion methods.
As a result, we find a new phase with an in-plane antiferromagnetism in 2D,
which appears by the similar mechanism as that of the pion condensate phase (so-called ``Aoki phase'') in lattice QCD with Wilson fermion \cite{Araki_2013}.
On the other hand, such a phase does not appear in 3D,
and the electron correlation results in the shifting of the topological phase boundary \cite{Sekine_2013}.

\section{2D quantum spin Hall insulator}
QSH effect is characterized by the quantized spin Hall conductivity $\sigma_{xy}^{s} = \sigma_{xy}^{\uparrow}-\sigma_{xy}^{\downarrow}=e/2\pi$.
The appearance of QSH state is led by spin-orbit interaction, which keeps the time-reversal symmetry but breaks the spin symmetry
so that it leaves odd numbers of Kramers pairs (pairs of states related by time inversion).
It was first suggested by Kane and Mel\'e by incorporating the spin-orbit interaction to the conventional tight-binding model of honeycomb lattice \cite{Kane-Mele},
and later realized in the HgTe quantum well \cite{Konig_2007}.
It is one of the symmetry-protected topological insulators, characterized by the topologically invariant ``spin Chern number'',
which takes a value either 1 (topological; QSH effect) or 0 (trivial; normal insulator).

\subsection{Kane--Mel\'e model}

Here we use the Kane--Mel\'e model to describe a QSH insulator,
which consists of the nearest-neighbor hopping term $H_T$ and the spin-orbit coupling term $H_{SO}$.
The first term $H_T = -t\sum_{\langle \bfr,\bfr' \rangle} \left[a^\dag(\bfr)b(\bfr') +\mathrm{H.c.}\right]$,
where $a$ and $b$ are the creation/annihilation operators of electrons on the sublattices A and B of the honeycomb lattice,
 is usually seen in the tight-binding model of graphene,
which reveals two independent zero-energy points (Dirac points) $\bfK_\pm$ in the Brillouin zone $\Omega$ \cite{Wallace_1947}.
The degeneracy of these ``valleys'' can be regarded as Fermion doublers according to the Nielsen--Ninomiya's theorem \cite{Nielsen-Ninomiya}.

The spin-orbit coupling term is defined as next-to nearest neighbor hopping with the amplitude $t'$,
which gives
$H_{SO} = - \sum_{\bfk \in \Omega} u(\bfk) \left[a^\dag(\bfk)\sigma_z a(\bfk) -b^\dag(\bfk)\sigma_z b(\bfk)\right]$
in the momentum space,
where $\sigma_z$ is the Pauli matrix acting on the spin components.
Since the kernel $u(\bfk)$ takes a finite value $\pm 3\sqrt{3}t'$ at each Dirac point $\bfK_\pm$,
it opens a finite bandgap $|3\sqrt{3}t'|$ at the Dirac points.
This term behaves as a momentum-dependent mass term, similar to the Wilson term in lattice fermion construction \cite{Wilson-fermion},
which splits the degeneracy between two valleys.

In addition to the ``Wilson term'',
here we also introduce the ``normal mass term'' $ H_M = m \sum \left[a^\dag \sigma_z a - b^\dag \sigma_z b \right]$,
namely a staggered Zeeman term,
which induces a uniform bandgap $m$ at each Dirac point.
Both $H_{SO}$ and $H_{M}$ breaks the sublattice exchange symmetry and spin SU(2) symmetry (down to U(1)),
but their momentum dependence are different.
If both of them are incorporated,
they yield an effective mass $m\pm3\sqrt{3}t'$ for each valley $\bfK_\pm$.
The topological phase structure of the system is characterized by the signs of these mass terms.
If they are opposite $(|3\sqrt{3}t'|>|m|)$, the system reveals the QSH effect;
while it becomes a normal insulator when they have the same sign $(|3\sqrt{3}t'|<|m|)$.
Transition between these phases occurs at $|m|=|3\sqrt{3}t'|$, accompanied with the gap closing at one of the valleys.
Our aim is to investigate how this topological phase structure evolves in the presence of the electron-electron interaction.

\subsection{Strong coupling analysis with lattice gauge theory description}
In order to apply the strong coupling expansion technique to analyze the effect of electron correlation,
here we construct the imaginary time lattice action from the Hamiltonian above.
Imaginary time direction is discretized by a finite lattice spacing $\Delta\tau$,
and we introduce there U(1) link variables $U_0 = e^{ie\Delta\tau A_0}$ to define the coupling to the electric field.
Since the time discretization generates new unnecessary doublers,
here we reduce the degrees of freedom by suppressing the spin indices and regarding the doublers as the spin degrees of freedom,
like in the staggered fermion formalism of lattice fermions.
As an artifact of the discretization, spin SU(2) is broken down to U(1), like the flavor (taste) symmetry breaking in the staggered fermion formalism \cite{Susskind_1977,Sharatchandra_1981}.
Here we regard this U(1) as the symmetry group within the $(\sigma_x,\sigma_z)$-plane generated by $\sigma_y$,
which we call ``remnant U(1) spin'' symmetry.
We mainly focus on the breaking of this remnant U(1) symmetry.

Dynamics of the gauge field is included in the action as the plaquette terms,
i.e. gauge invariant polynomials of link variables.
As this term becomes proportional to the inverse coupling strength parameter $\beta=\epsilon_0 \vf/e^2$,
we can perform the perturbative expansion by $\beta$ in the strong coupling region.
Here we neglect this term to focus on the strong coupling limit for simplicity.

In the strong coupling limit $(\beta=0)$, we can integrate out the gauge field,
leading to the on-site density-density interaction term connecting the different time points,
which is similar to the on-site repulsion term in the phenomenological Hubbard model.
In order to treat this term analytically,
we take a mean-field ansatz with a complex order parameter $\sigma=\sigma_1+i\sigma_2$ respecting the remnant U(1) symmetry,
like the chiral and pion condensates in QCD ($\sigma_1$ and $\sigma_2$ should not be confused with the Pauli matrices $\sigma_{x,y,z}$).
Its real and imaginary parts correspond to the order parameter for antiferromagnetism (i.e. two sublattices are spin-polarized in opposite directions)
in $z$-direction and $x$-direction respectively.
We should note that both the spin-orbit coupling $t'$ and the staggered Zeeman field $m$ break the remnant U(1) symmetry explicitly,
in the $\sigma_1$-direction.
Integrating out the fermionic fields,
we obtain the thermodynamic potential (free energy) $F_\mathrm{eff}(\sigma)$.
Solving the gap equations
$\partial F_\mathrm{eff}/\partial \sigma_{1,2} |_{(\tilde{\sigma}_1,\tilde{\sigma}_2)} =0$,
we search for the potential minimum $(\tilde{\sigma}_1,\tilde{\sigma}_2)$ as a function of the spin-orbit coupling $t'$ and the staggered magnetic field $m$
(see Ref.\cite{Araki_2013} for the details of calculation).

\subsection{Phase structure}
\begin{figure}[tbp]
\begin{tabular}[b]{cc}
 \includegraphics[width=7cm]{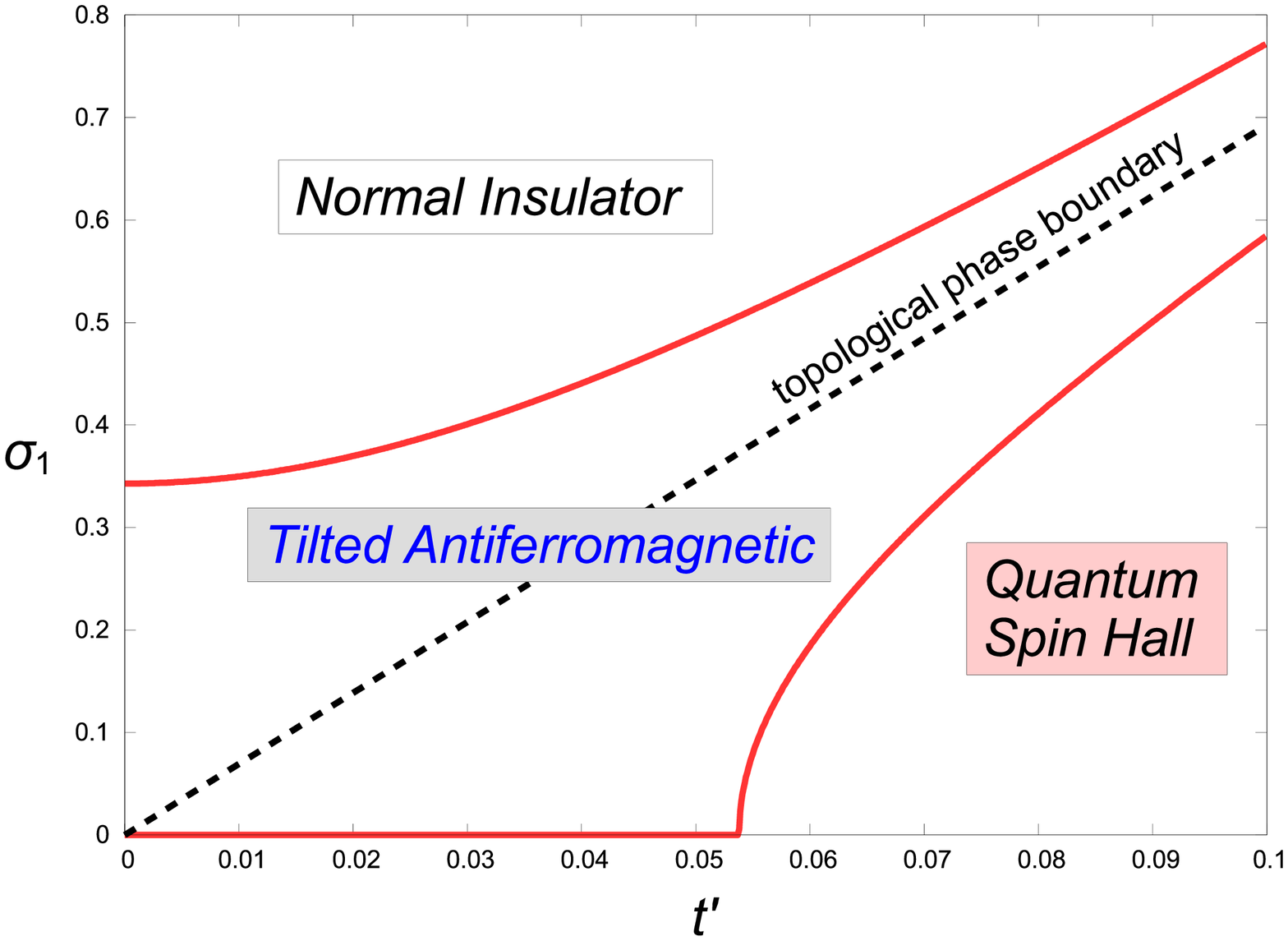}
&
{\footnotesize
\begin{tabular}[b]{c|c}
\hline
 \textbf{Kane--Mel\'e model} & \textbf{Lattice QCD} \\
\hline \hline
 staggered Zeeman field $(m)$ & mass term \\
\hline
 spin-orbit interaction $(t')$ & Wilson term \\
\hline
 valleys & doublers \\
\hline
 remnant U(1) spin symmetry & continuous chiral symmetry \\
\hline
 photons (electromagnetic field) & gluons \\
\hline
 Tilted AF phase & Aoki phase \\
\hline
 $\langle \sigma_2 \rangle (\sim \langle a^\dag \sigma_x a - b^\dag \sigma_x b\rangle)$ & $\langle \bar{\psi} i\gamma_5 \psi \rangle$ (pion condensation) \\
\hline
\multicolumn{2}{c}{\vspace{1cm}} \\
\end{tabular}
}
\end{tabular}
\vspace{-25pt}
\caption{The phase diagram in the $(t',\sigma_1)$-space.
There appears a new ``tilted antiferromagnetic (TAF)'' phase ($\sigma_2 \neq 0$),
between the normal insulator (NI) phase ($\sigma_2=0$ and $\sigma_1/2 > 3\sqrt{3}t'$)
and the quantum spin Hall (QSH) phase ($\sigma_2=0$ and $\sigma_1/2 > 3\sqrt{3}t'$).
The table on the right hand side shows the correspondence between this phase structure and the phase structure of lattice QCD with Wilson fermion.}
\label{fig:phasediagram}
\end{figure}

Now we can observe the phase structure of the interacting system,
characterized by the presence or absence of the order parameter $\sigma_2$,
which represents the antiferromagnetic order orthogonal to the spin-orbit interaction $t'$ and the staggered magnetic field $m$.
Since $\sigma_1$, which serves as the renormalized mass term $m$, is a monotonically increasing function of $m$,
here we use $\sigma_1$ as a parameter in the phase diagram instead of $m$.
The phase diagram can be separated into three phases as shown in Fig.\ref{fig:phasediagram}.
Two of them, { Quantum spin Hall (QSH)} phase and { normal insulator (NI)} phase,
have the same characteristics as those appearing in the noninteracting system.
The antiferromagnetism is aligned in the $z$-direction (i.e. $\sigma_2=0$),
and they are characterized whether they are above or below the topological phase boundary $\sigma_1 = 3\sqrt{3}t'$.

In contrast to the phase structure of noninteracting system (shown by the dotted line in Fig.\ref{fig:phasediagram}),
there evolves a new { tilted antiferromagnetic (TAF)} phase between QSH and NI phases.
Here the antiferromagnetic order is tilted towards the in-plane direction, i.e. $\sigma_2 \neq 0$.
In the absence of the staggered magnetic field $m$,
$\sigma$ is completely tilted to the $\sigma_2$-direction,
which is consistent with the ``XY-antiferromagnetic insulator'' phase
found in the analysis of the Kane--Mel\'e--Hubbard model \cite{Muramatsu_Assaad,Reuther,Sekine_2012}.
Going under the TAF phase, the phase transition between the QSH phase and the NI phase
occurs without closing the bandgap of fermions.
This behavior appears to contradict the previous studies in the noninteracting Dirac fermion system,
where the gap closing is essential for the topological phase transition \cite{Murakami_2007}.
In this system, however, the QSH phase is justified not by the time-reversal symmetry but by the spin rotation symmetry by $\sigma_z$.
Therefore, transition to the normal state without gap closing is allowed by breaking the symmetry $\sigma_z$,
by the in-plane antiferromagnetism in the TAF phase.
Recently it has been proposed that this phase structure can be characterized by a new quantity,
called symmetry-protected topological charge, which takes a non-integer value in the $\sigma_z$-broken phase \cite{Ezawa_2013}.


We can interpret the phase structure obtained here in analogy with the phase structure of lattice QCD with Wilson fermions.
Since both the spin-orbit coupling term in the Kane--Mel\'e model and the Wilson term in the lattice fermion formalism break the degeneracy of valleys (doublers).
In the strong coupling region of the ordinary QCD, the chiral symmetry,
the continuous symmetry generated by the Dirac matrix $\gamma_5$ (exchange of left-handed and right-handed fermions), gets spontaneously broken,
characterized by the chiral condensate $\langle \bar{\psi}\psi \rangle$ as the order parameter.
In the presence of the Wilson term, there appears a new phase called ``Aoki phase'' \cite{Aoki}.
It is characterized by the pion condensate $\langle \bar{\psi}i\gamma_5\psi \rangle$ as the order parameter,
which is orthogonal to the chiral condensate explicitly pointed by the mass term and the Wilson term.
This situation is similar to that of the in-plane antiferromagnetism $\sigma_2$ in our calculation,
which is orthogonal to the perpendicular antiferromagnetism $\sigma_1$, explicitly pointed by the spin-orbit interaction $t'$ and the staggered Zeeman field $m$
(their correspondence is summarized in the table in Fig.\ref{fig:phasediagram}).
Thus we can consider that the TAF phase here appears by the similar mechanism as that of Aoki phase in lattice QCD.

\section{3D topological insulator}
3D topological insulator is characterized by the existence of gapless surface states, protected by time-reversal symmetry.
It is ensured by the strong spin-orbit coupling,
which changes the parity of the system by the level crossing of the highest occupied state and the lowest unoccupied state,
realized in the materials such as $\mathrm{Bi}_2 \mathrm{Se}_3$  \cite{Zhang_2009}.
The 3D bulk state can be described by the Wilson fermion, $H_0(\bfk) = \sum_j (\sin k_j)\alpha_j + m(\bfk)\beta$,
where $\alpha_{j}=i\gamma_0\gamma_j \ (j=1,2,3)$ and $\beta=\gamma_0$ are Dirac matrices.
The basis is defined as $(c_{A\uparrow}^\dag,c_{A\downarrow}^\dag,c_{B\uparrow}^\dag,c_{B\downarrow}^\dag)$,
where $c_{X\sigma}^\dag$ is a creation operator of an electron at orbital (pseudospin) $X(=A,B)$ with spin $\sigma(=\uparrow,\downarrow)$.
The mass term has momentum dependence, $m(\bfk) = m_0 + r\sum_j (1-\cos k_j)$,
where the ``Wilson term'' $r$ corresponds to the amplitude of the spin-orbit coupling.
The system becomes a strong topological insulator, which is characterized by odd number of gapless surface states,
for $-2r <m_0 <0$.

Here we investigate how this topological phase structure is altered by the electron-electron interaction.
Similar to the 2D case, we construct the Euclidean lattice action from the Hamiltonian,
and introduce the interaction as the coupling to the electric field, in terms of U(1) link variables $U_0$ in the temporal direction.
We perform the strong coupling expansion by the inverse coupling strength $\beta = \epsilon \vf/e^2$,
and integrate out the link variables, leading to the 4-Fermi coupling terms.
We introduce two types of bosonic mean fields,
$\phi_\sigma \equiv \langle \bar{\psi}\psi\rangle$ and $\phi_\sigma \equiv \langle \bar{\psi}i\gamma_5\psi\rangle$,
which correspond to the pseudospin antiferromagnetism in two directions, orthogonal to each other.

\begin{figure}[tbp]
\begin{center}
 \includegraphics[width=6cm]{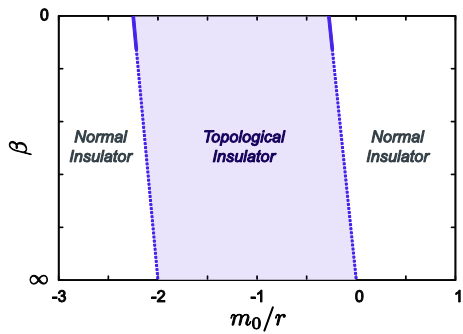}
\end{center}
\vspace{-20pt}
\caption{A possible phase diagram of the 3D topological insulator with the ``Wilson term'' $r=0.5$.
It is based on the mean-field calculation around the strong coupling limit $(\beta=0)$,
and extrapolated to the noninteracting limit $(\beta=\infty)$.}
\label{fig:3d}
\end{figure}

We find the expectation values of the order parameters by minimizing the effective potential obtained by integrating out the fermionic degrees of freedom
(see Ref.\cite{Sekine_2013} for the details of calculation).
Here we should note that $\phi_\pi$ vanishes even in the strong coupling limit $\beta=0$, unlike in the 2D case,
since the momentum integral at the one-loop level converges in $(3+1)$-dimensions.
Thus an ``Aoki phase'' does not appear in between the topological insulator (TI) and the normal insulator (NI),
and the effect of the interaction results in the shift of the boundary between these two phases,
characterized by the effective mass $m_\mathrm{eff} = m_0+(1-r^2)\phi_\sigma/2$.
The phase structure under the electron-electron interaction, extrapolated to the noninteracting limit $(\beta=\infty)$,
is shown in Fig.\ref{fig:3d}.

\section{Conclusion}
In this paper, we have investigated the effect of electron correlation on the phase structure of 2D and 3D topological insulators,
by using the idea of strong coupling expansion of lattice gauge theory.
We have shown that a new phase with the in-plane antiferromagnetic order appears in 2D topological insulator,
similar to the ``Aoki phase'' in lattice QCD with Wilson fermion,
while such a phase does not emerge in the 3D case due to the lattice regularization of the interaction strength.
Application of our methods to more complex systems,
such as the diamond lattice and the anisotropic topological insulators, would be a challenging problem.
Recently it has been proposed that the phase transition between topological (quantum Hall) state and normal state
can also be driven by the random disorder potential \cite{Beenakker}.
It would be interesting to observe the shift of such a topological phase structure under the symmetry-breaking disorders,
such as magnetic impurities,
to see whether a new phase like the ``Aoki phase'' in our phase diagram can appear in the midst of the topological phase transition.

\subsection*{Acknowledgments}
Y.~A. is thankful to T.~Hatsuda, T.~Oka, and N.~Tsuji for fruitful discussions.
Y.~A. is supported by JSPS Postdoctoral Fellowship for Research Abroad (No.25-56).
T.~K. is supported by JSPS Research Fellowship for Young Scientists (No.23-4302).
This work is partially supported by the Grants-in-Aid for Scientific Research (No. 24740211 and
No. 25103703) from the Ministry of Education, Culture, Sports, Science and Technology, Japan (MEXT).

\end{document}